\documentclass{article}
\usepackage[utf8]{inputenc}
\usepackage{graphicx}
\usepackage[sorting=none]{biblatex}
\usepackage{amsmath}
\usepackage[margin=1.2in]{geometry}
\usepackage{multicol}
\usepackage{float}
\usepackage{xcolor}
\usepackage{physics}
\usepackage{authblk}

\title{Matrix Product State Pre-Training for Quantum Machine Learning}
\author[1]{James Dborin}
\author[1,2]{Fergus Barratt} 
\author[1]{Vinul Wimalaweera}
\author[2]{Lewis Wright}
\author[1]{Andrew G. Green}

\affil[1]{\small \emph{ London Centre for Nanotechnology, University College London, London, United Kingdom}}
\affil[2]{\small \emph{Department of Mathematics, King’s College London, London, United Kingdom}}

\date{}

\begin{document}

\maketitle

\begin{abstract}
        Hybrid Quantum-Classical algorithms are a promising candidate for developing uses for NISQ devices. In particular, Parametrised Quantum Circuits (PQCs) paired with classical optimizers have been used as a basis for quantum chemistry and quantum optimization problems. Training PQCs relies on methods to overcome the fact that the gradients of PQCs vanish exponentially in the size of the circuits used. Tensor network methods are being increasingly used as a classical machine learning tool, as well as a tool for studying quantum systems. We introduce a circuit pre-training method based on matrix product state machine learning methods, and demonstrate that it accelerates training of PQCs for both supervised learning, energy minimization, and combinatorial optimization.

\end{abstract}

\begin{multicols}{2}
\section{Introduction}

    Parametrised Quantum Circuits (PQCs) have been the focus of attempts to demonstrate quantum computational advantage on NISQ devices, for problems of both scientific and commercial interest. Typically these efforts involve parametrising a quantum circuit with a series of rotation angles, and using a classical optimizer to find a set of angles which minimizes a given cost function. The quantum device is used to estimate the cost function associated with a particular set of parameters, where it is assumed to be hard to calculate the cost function classically. Algorithms based on these methods are often called hybrid quantum-classical algorithms.
    
    A major hurdle in developing useful hybrid algorithms is the problem of vanishing gradients. It has been shown that the size of initial gradients decrease exponentially towards zero as the number of qubits and the depth of the circuits increases, when parameters are randomly initialised \cite{BarrenPlateau}. Other circuit metrics have been demonstrated to produce these so-called barren plateaus, such as ansatz expressibility \cite{ansatz_plateau}, the entanglement between hidden and visible nodes in the circuit \cite{EntanglementPlateaus, entanglementMitigation}, and circuit noise~\cite{noise_barren_plateau}. 
    
    The existence of the barren plateau has motivated attempts to improve PQC learning algorithms to avoid training costs growing exponentially. These include developing gradient free algorithms~\cite{rotosolve, analytic_gradient}, defining local cost functions as targets for learning algorithms~\cite{cost_function_dependant_plateaus}, and initialisation schemes~\cite{block_iden}.
    
    Here we introduce a novel initialisation scheme based on tensor network algorithms. Tensor network based methods are the state of the art for numerical simulations of 1D and 2D spin systems~\cite{OrusTN}, and also for the simulation of quantum circuits~\cite{TensorNetwork}. Tensor networks have been used to solve optimization problems such as portofolio optimization, and are found to be competitive with commercial solvers~\cite{PortfolioOptTN,QEnhancedOpt}.  Recently, tensor network based methods have also been used as the basis for machine learning algorithms. Matrix Product States (MPS) have been trained as a classifier for several machine learning tasks \cite{SupervisedMPSLearning, Unsupervised_Modelling_Samples}. The 2D generalisation of MPS, Projected Entangled Pair States (PEPS), have also been used for image classification\cite{PEPS_CNN}. Both Tree Tensor Networks (TTN) and MERA networks have also been used as image classifiers \cite{towards_qml, HierarchicalQC}.

\end{multicols}
\newpage

\begin{figure}[t]
    \includegraphics[width=\linewidth]{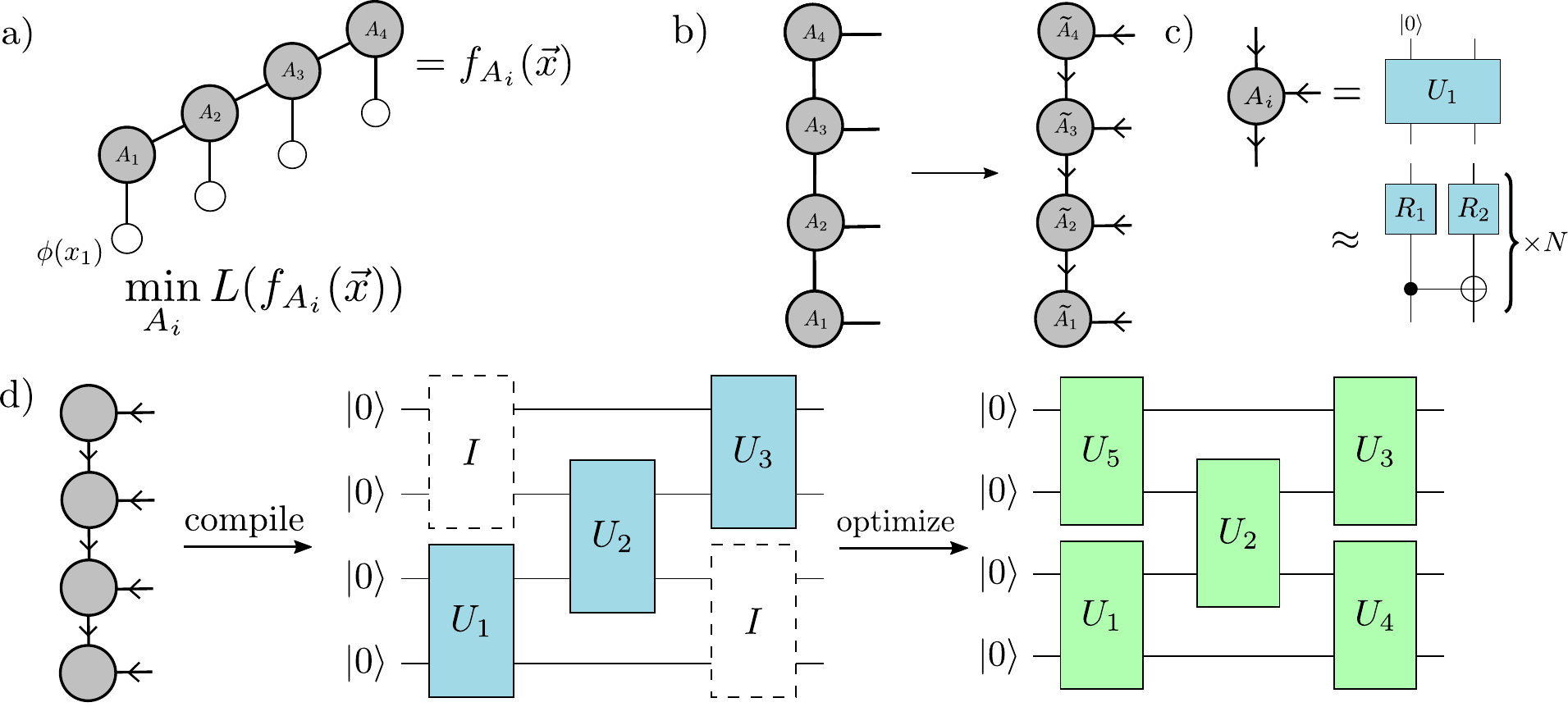}
    \caption{\textbf{Outline of the MPS initialisation procedure}. a) \emph{Classical Training -} A MPS is optimized as an approximate solution to a given problem instance, by variationally minimizing a cost function over the tensors, $A_{i}$. Classical data points, $x_{i}$, are lifted into a higher dimensional Hilbert space by way of a mapping $\phi(x_{i}) = ( \cos(x_{i}), \sin(x_{i}) )$. b) \emph{Canonicalise -} The minimized MPS is put into canonical form, where the tensors are isometries from the space of incoming arrows to the space of out going arrows. Any MPS can be put in this form, due to a gauge freedom in the MPS\cite{mps_representations}. c) \emph{Embed and Compile -} The isometric tensors are exactly embedded into unitary matrices. This can be further decomposed approximately using gates available on the quantum device. d) \emph{Quantum Training -} The compiled unitaries are placed in a circuit along the diagonals, and the other gates in the circuit are initialised to the identity. This ensures that this circuit is as good an approximation to the optimal solution as the MPS is after the compilation process. The optimizer is then able to vary all the parameters, including those in the gates initialised to the identity. }
    \label{fig:AlgorithmOutline}
\end{figure}

\begin{multicols}{2}

    In this work, we optimize tensor networks as candidate ground states, as classifiers, and as solutions to combinatorial optimization problems. We use these networks to seed a PQC with an effective set of starting parameters, before continuing to train the quantum circuit. By beginning training in a part of the parameter space that is close to the target state, the number of steps to reach the desired minima can be reduced. In this way the impact of the barren plateau is reduced and limited quantum resources can be used more effectively during training. From here onwards this procedure will be known as \emph{MPS Pretraining}.

\section{MPS Pretraining}
    
    MPS pretraining involves three steps:
    \begin{itemize}
        \item[1.] Train a tensor network 
        \item[2.] Compile the tensor network into gates
        \item[3.] Initialise a circuit with these gates
    \end{itemize}
    Fig.~\ref{fig:AlgorithmOutline} outlines this procedure.
    
    \subsection{MPS Optimization}
    
    Given a problem instance, such as a Hamiltonian or a labeled data set, and a cost function where the minimum corresponds to the state of interest, the first step is to produce a MPS that minimizes the given cost function. The bond dimension of a MPS, denoted $\chi$, is a parameter that determines the complexity of the MPS model. Higher bond dimension models tend to produce better results, at the cost of greater training complexity.  This training is done purely classically, and software exists both to efficiently contract and optimize these states, for example we refer the reader to Refs.~ \cite{itensor,TensorNetwork, tenpy, quimb}. In the following we discuss MPS algorithms, but the insights apply to other tensor network architectures.
    
    There are a variety of algorithms that can be used to optimize MPS. To find the ground states of Hamiltonians the DMRG algorithm 
    is known to be very effective~\cite{SchollwoeckDMRG}. Similar optimisation algorithms include imaginary-time TEBD and TDVP, both of which are used in this work~\cite{TDVP,TEBD}. In Ref.~\cite{SupervisedMPSLearning} a DMRG-inspired machine learning algorithm is introduced which we adapt for this work. Note that in principle gradient descent can always be performed directly on all the tensors in the tensor network. This more closely reflects classical machine learning algorithms. However these quantum-inspired learning algorithms tend to perform as well in practice, and often perform better for quantum-based problems such as energy minimisation. 
    
    \subsection{MPS Compilation}
    
    Having trained a MPS to minimize a cost function, the next step is to represent the MPS as a set of rotation angles in a PQC. As opposed to other classical machine learning methods, low bond dimension tensor networks permit efficient representations on quantum circuits \cite{QMPS}. MPS have a gauge freedom which means that any MPS can be brought into \emph{canonical form} in which each tensor is an isometry~\cite{mps_representations}. These can then easily be embedded in unitary matrices. Bond dimension 2 MPS can be expressed as a staircase of 2 qubit unitary gates~\cite{QMPS}. The rotation angles are extracted from these unitaries with some compilation scheme. For 2 qubit unitaries we use the KAK Decomposition \cite{KAK_Decomp}, which decomposes 2 qubit unitaries into 4 single qubit gates with 3 rotation angles each, and 3 different 2-qubit interaction gates. If necessary these gates can be further decomposed into hardware efficient gates. Restricted gate sets can be used as compilation targets, and the rotation angles can be found variationally~\cite{QuantumCompiling}. Later we suggest an approximate compilation scheme, based on insights from Ref.~\cite{reverse_stair_ansatz}, which ensures that the largest unitary matrices that need to be compiled are 2-qubit gates.
    
    Often brick wall circuits are used as ans\:atze for PQC research, being the most general way to parametrise a quantum circuit limited to nearest neighbour interactions. To initialise a brick wall circuit with a pretrained MPS, the gates on the diagonal of the brick wall must be initialised with the angles extracted in the compilation of the MPS. All of the off-diagonal gates are initialised to the identity. This ensures that before quantum training begins, the circuit represents exactly as good a candidate for the target state as the MPS is after compilation is completed. The optimizer is then free to vary the angles of all of the gates. We demonstrate that starting with this initialisation accelerates training, requires fewer gradient updates, and avoids local minima.

\section{Results}

\subsection{Combinatorial Optimization}\label{sec:qaoa}

    We test the initialisation scheme on the Max Cut optimisation problem which is often used as a benchmark for QAOA algorithms~\cite{qaoa_farhi}. In the Max Cut problem a graph, $G(E,V)$, is provided along with weights, $w_{ij}$, on each edge. The task is to find a set of vertices, $S$, such that the total weight of the edges connecting $S$ to their complement is maximized. This is equivalent to finding the ground state of the Hamiltonian,
    
    \begin{equation}
        H = \sum_{\langle i,j\rangle} w_{ij}(1 - Z_{i}Z_{j})
    \end{equation}
    where $Z$ is the Pauli Z matrix. 
    
\begin{figure}[H]
    \centering
    \includegraphics[width=\linewidth]{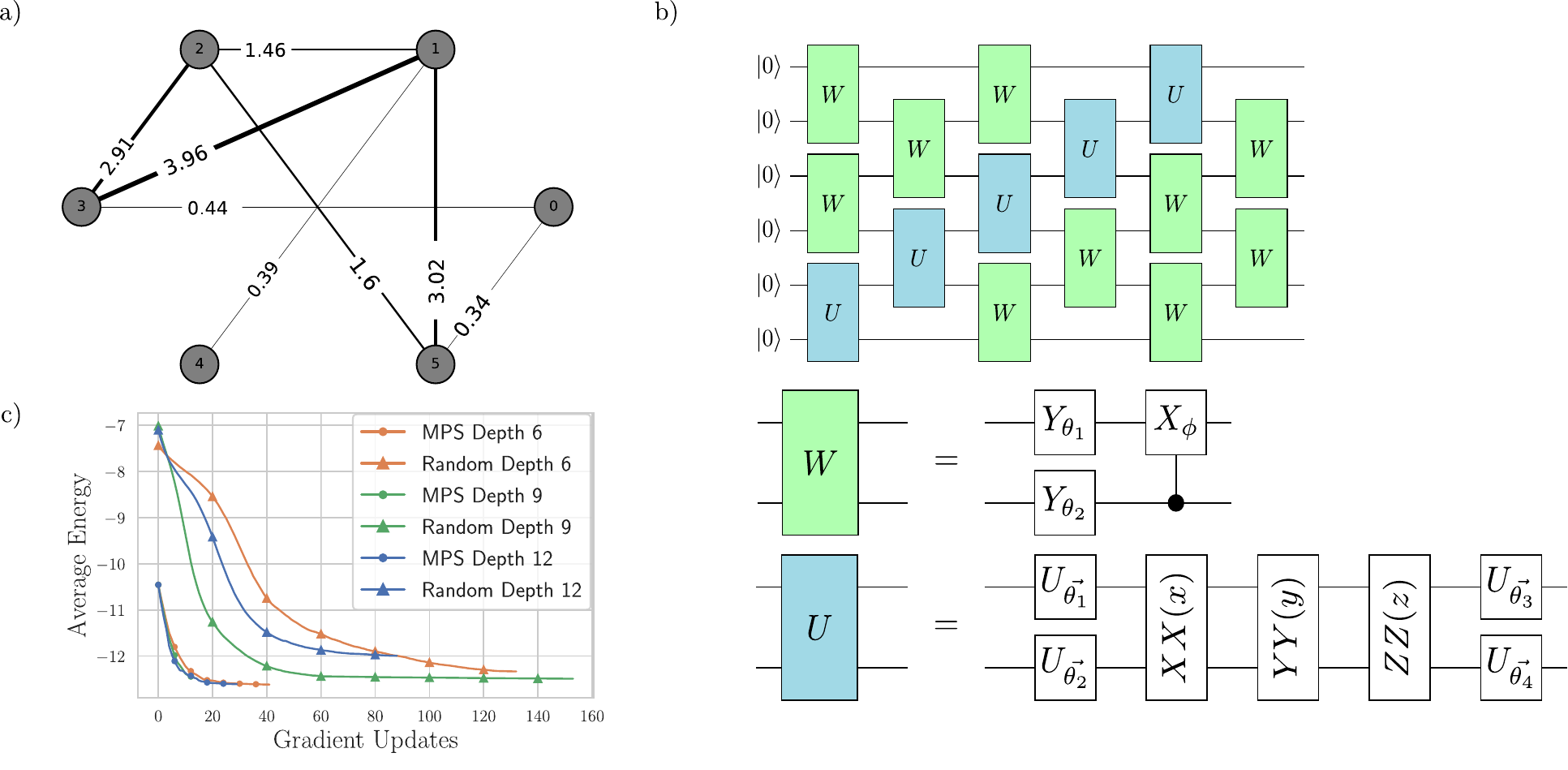}
    \caption{\textbf{Max Cut optimization} a) \emph{Max Cut Graph - } The 6 node max cut problem graph with the edge weights. The Max Cut Hamiltonian is given as $H = \sum_{\langle i,j\rangle} w_{ij}(1 - Z_{i}Z_{j})$, where $w_{ij}$ is the weight along the edge connecting node $i$ to node $j$. Each node is encoded as a qubit. b)\emph{Quantum Circuit Ansatz -} For all circuits the diagonal unitaries were expressed fully using the KAK decomposition\cite{KAK_Decomp}, whereas off diagonal unitary matrices were made with combinations of Y rotations and controlled X rotations. Circuits were optimized using gradient descent with decaying learning rate. c) \emph{MPS Pretraining results -} Compared to randomly initialised circuits, of depth 6, 9, and 12, the circuits initialised with MPS-compiled angles converge with fewer gradient steps, and converge to a better ground state estimate than than the randomly initialised counterparts. In the MPS initialised circuits there is almost no change in performance as depth increases, whereas increasing depth with randomly initialised parameters slows down optimization. }
    \label{fig:QAOARes}
\end{figure}

    We use imaginary time TEBD to find a bond dimension 2 approximation to the ground state for an instance of the Max Cut problem on a graph with 6 vertices. In Fig.~\ref{fig:QAOARes} we show the results of training with this initialisation for depth 6, 9, and 12 circuits. Optimization was performed for an imaginary time of $3\times 10^{-2}$, with a time step of $1\times 10^{-3}$. We only optimize the MPS for short periods of time because for small problem instances low bond dimension MPS can often get close to the optimal answer. To more closely match the performance on large problem instances which require quantum circuits that cannot be easily simulated we do not fully optimize the MPS.

\end{multicols}
\newpage

\begin{multicols}{2}

 We compare training starting from MPS pretrained circuits to that starting from random initialisations of the same ansatz. Instead of the standard QAOA scheme, we use a brick wall circuit made up of independent 2-qubit gates. These sorts of circuits have been explored for optimisation problems and their performance is competitive with QAOA \cite{VQEvQAOA}. 
    
    The MPS initialised circuits start off at a better energy than the random counterparts, which is to be expected. 
    However the initial steps from the MPS initialised states are noticeably larger than those taken from randomly initialised circuits. 
    These circuits reach a minimum with fewer gradient descent updates, and reach better minima than the any of the randomly initialised states. 
    In the random state there is a notable decrease in performance between the depth 9 and depth 12 ansatz, where the depth 12 gets stuck in a local minimum. 
    No drop in performance is observed in the MPS initialised circuits. 
    
    \subsection{Finding Ground States}
    
    We also implement MPS pretraining for the purpose of finding ground states of electronic Hamiltonians, a common benchmark in hybrid quantum classical algorithm research. We use imaginary time TDVP to estimate the ground state of the electronic Hamiltonian of $H_{2}$ and $LiH$. The Hamiltonians are 
    constructed using the OpenFermion package~\cite{openFermion}. The same is done for the Transverse Field Ising Model (TFIM) which is a widely studied Hamiltonian in condensed matter physics. 
    
\begin{figure}[H]
    \centering
    \includegraphics[width=0.8\linewidth]{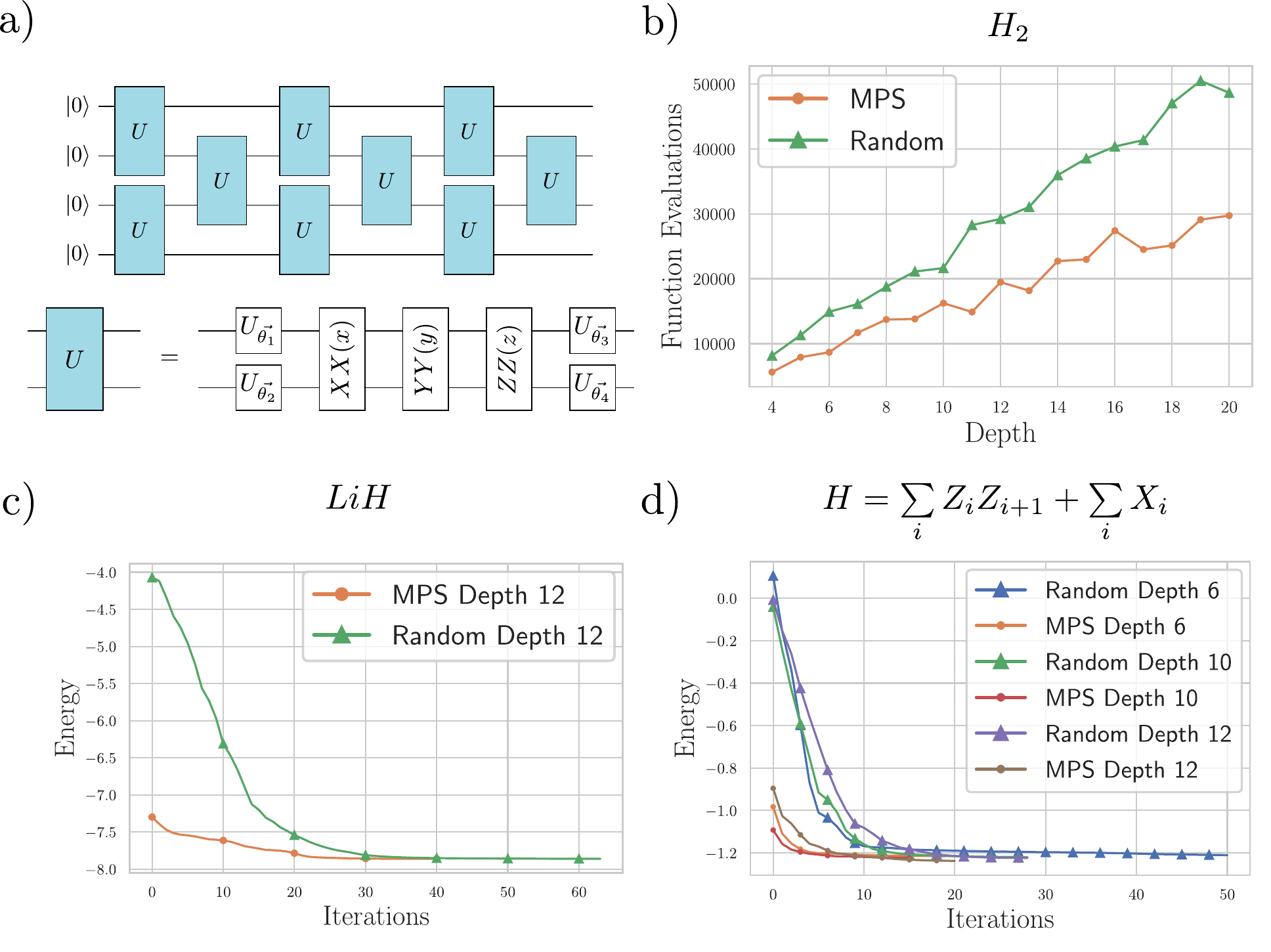}
    \caption{\textbf{Initialising VQE with MPS} a) \emph{Quantum Circuit Ansatz - }The structure of the circuits used for these experiments. Shown is a depth 6 circuit on 4 qubits. Each unitary, $U$, is decomposed into 15 rotation angles using the KAK Decomposition. The single qubit gates are made up of three independent rotations, Z followed by Y followed by Z. All the angles in the circuit are independent. All optimization was performed using the BFGS algorithm. b) \emph{$H_{2}$ Optimisation -} For all depths the circuits produced a good estimate for the ground state energy, $E-E_{min}<1\times 10^{-8}$. The MPS initialised circuits required thousands of fewer function evaluations at each depth, and the number of iterations is growing more slowly with depth than for the randomly initialised circuits.  c) \emph{LiH Optimisation -} For the depth tested approximately half as many function evaluations are required to optimize the cost function when using MPS initialisation. The initial gradients are larger in the MPS case, and the optimization begins closer to the minima. d) \emph{TFIM Optimisation - } At all depths tested the MPS initialisation converged after fewer iterations, and to lower ground state energies.}
    \label{fig:VQEH2}
\end{figure}

    In all these cases imaginary time TDVP was used to construct a MPS approximation to the ground state, and this MPS is used to initialise a quantum circuit. For the $H_2$ and $LiH$ problems, the MPS converges to the same ground state as the optimized quantum circuit, suggesting either low bond dimension MPS are good approximations to these ground states, or the brick wall circuits do not offer significantly more expressivity in this problem.  To demonstrate that even in cases where low bond dimension MPS are not as effective, we run the TDVP algorithm for a shorter time, meaning a worse approximation is used to initialise the circuits. These results are given in Fig.~\ref{fig:VQEH2}.
    
    In all three cases we find significant decreases in the number of iterations needed to reach the lowest energy states, compared to the randomly initialised circuits. We also compared circuits initialised to the identity, with all rotation angles set to 0. In each case the zero initialised circuit failed to optimized, and quickly ended up in a local minima.

    Fig.~\ref{fig:VQEH2}b compares the performance of the MPS pretrained circuit with a randomly initialised circuit in finding the ground state of the $H_{2}$ electronic Hamiltonian. Optimisation of the MPS initialised circuit required fewer function evaluations than the randomly initialised circuits, while both found effectively the same ground state, differing by less that $1\times 10^{-8}$ at depth 10. The number of function evaluations needed for the MPS initialised circuit grows more slowly as a function of depth than for the randomly initialised circuit, implying that even relatively poor approximations of the ground state still initialise close to the target state. We see similar results for the LiH and TFIM Hamiltonians, Fig.~\ref{fig:VQEH2}c and d resprectively.

\end{multicols}
\begin{multicols}{2}

\newpage

\subsection{Machine Learning}
    Finally we pretrain quantum circuits to classify clothing labels using the Fashion MNIST dataset. The images were compressed so small circuits could be used to classify each image. This compression is done using principle component analysis on the training data set, and projecting the training images onto the principle components.
    The set of training images is collected into a matrix, $\bf{X}$. We compute the covariance matrix of the training dataset, given by
    
    \begin{equation}
        \bf{\Sigma} = \bf{X}^{T}\bf{X}
    \end{equation}
    
    Then we perform an SVD on the matrix $\bf{\Sigma}$
    
    \begin{equation}
        \bf{\Sigma} = \bf{U}\bf{\Lambda}\bf{U^{\dagger}}
    \end{equation}
    
    The principle components are identified as the columns of $\bf{U}$. To compress an image so that an $N$ qubit circuit can be used to classify the image we take the $N$ top principle components, $\{\vec{u}_{1},\vec{u}_{2},\cdots ,\vec{u}_{N}\}$, and take the inner product between each image (reshaped into a vector) and the principle component. The input to the $i^{th}$ qubit from the $j^{th}$ image in the training set is given by
    
    \begin{equation}
        \tilde{x}_{i,j} = \langle\vec{x}_{j},\vec{u}_{i}\rangle
    \end{equation}

\begin{figure}[H]
        \centering
        \includegraphics[width=\linewidth]{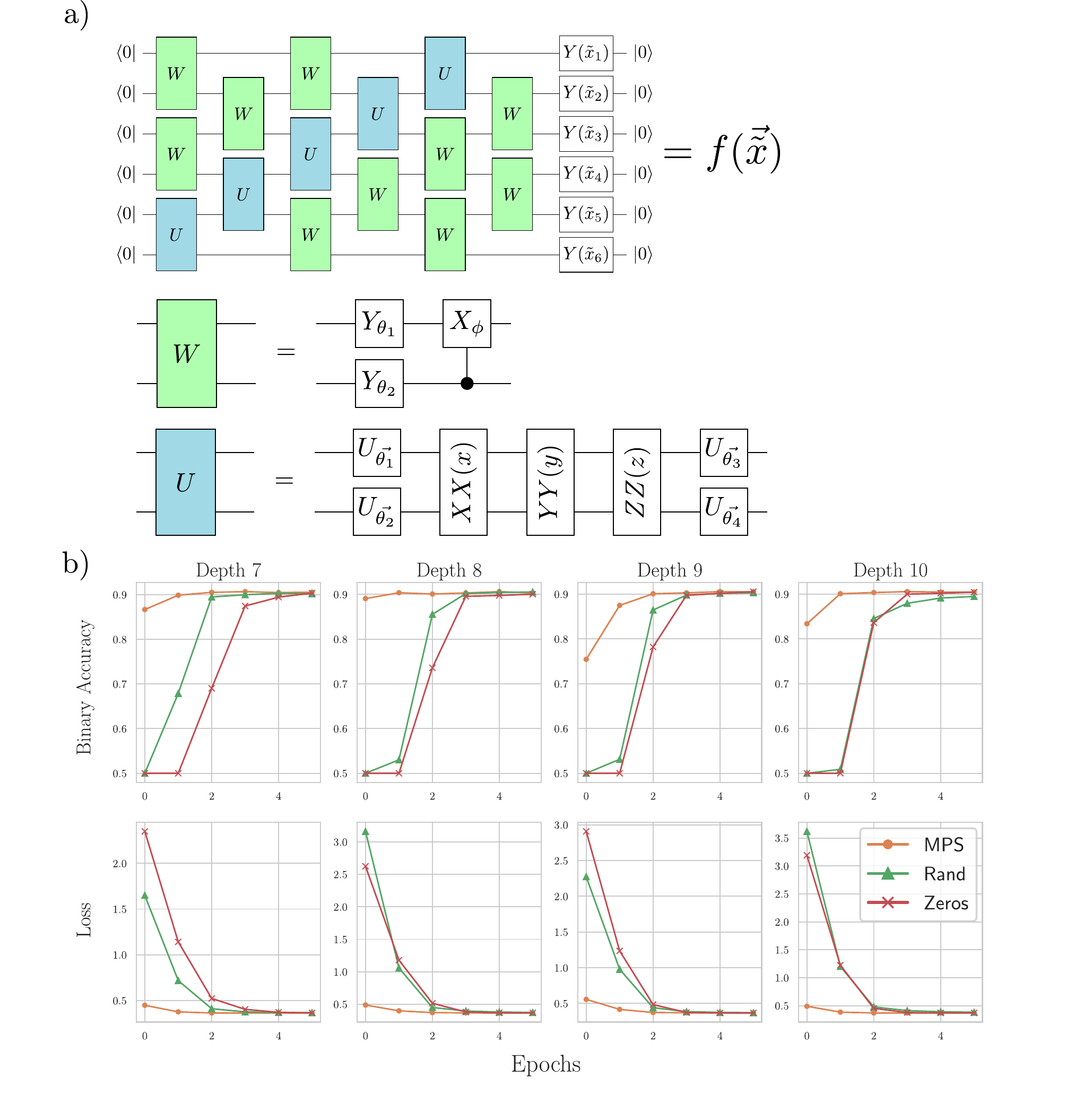}
        \caption{\textbf{Fashion MNIST Results.} a) \emph{Quantum Circuit Ansatz - } The circuit used in these classification experiments. U and W are defined in the same way as in Fig.~\ref{fig:QAOARes}a. The classifier ansatz is given by a brick wall circuit, and the data is introduced using Y rotations, parametrised by angles $\vec{\tilde{x}}$. A single component of this vector, $\tilde{x}_{i}$, is given by the projection of a single image onto the $i^{th}$ principle component of the training dataset. A decision function is defined as the probability of measuring the all-zero bit string. We use binary cross-entropy loss defined on the training dataset, $\tilde{\bf{X}}$, as a target to optimize the decision boundary. The circuit classifies the image with label 0 if $f(\vec{\tilde{x}}) < 0.5$ and 1 otherwise. Optimization is performed using the \emph{adam} optimizer. b) \emph{Results with MPS Initialisation -} The loss and binary accuracy on the training set during training is given as a function of the number of epochs trained for. An epoch is defined as the number of batched gradient updates required such that the entire training set has been processed once. In each case the number of qubits is the same as the depth. The MPS was trained for 5 epochs before being compiled to a quantum circuit and training is continued. The MPS initialised circuits took no more than 2 epochs to optimize, whereas the other initialisation schemes required 4 or more. As well as starting at a higher binary accuracy, the binary accuracy of the MPS initialised circuits also increased faster during early epochs, indicating that the local loss landscape of the MPS initialised circuits may be more easy to optimize over.}
        \label{fig:MNIST}
    \end{figure}

    Each projection was used as the input to a single qubit, so to use $N$ qubits, we projected an image onto the $N$ most significant principle components. The individual $\tilde{x}_{i}$ values are used as rotation angles in parametrised Y gates, Fig.~\ref{fig:MNIST}a. This method, as opposed to other compression methods, such as pooling, gives greater flexibility in the number of qubits that can be used. 
    The circuits used here were trained as binary classifiers. They were trained to distinguish between t-shirts and trousers. In Fig.~\ref{fig:MNIST}b we show the training set loss and accuracy at each epoch during training.     
    
    The MPS pretrained circuit is compared to both a random initialisation and an initialisation such that the circuit evaluates to the identity matrix. Ref.~\cite{block_iden} suggests the identity initialisation increases the size of initial gradients and is commonly used in quantum chemistry research. In this case all angles were set to zero, as all gates used were parameterised rotations. For an ansatz with static gates, more care must be taken so that the entire circuit evaluates to the identity.

    The MPS initialised circuit has a lower loss and higher accuracy on the training set during training than either the identity initialisation or the random initialisation. 

\end{multicols}
\newpage
    
\begin{multicols}{2}

\section{Compilation}

    For situations where bond dimension 2 MPS ansatz are not able to optimize a cost function, higher bond dimensions need to be used to capture a wider range of quantum states. Higher bond dimension MPS tensors are compiled to unitaries over more than 2 qubits, each doubling of the bond dimension requires another qubit in the representation of each tensor. This raises the problem of compiling high bond dimension MPS.
    
\begin{figure}[H]
    a)
    
    {\centering
    \includegraphics[width=\linewidth]{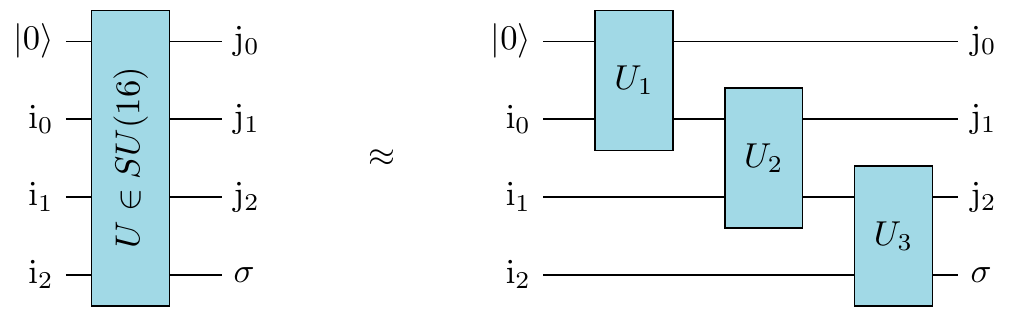}}
    b)
    
    {\centering
    \includegraphics[width=\linewidth]{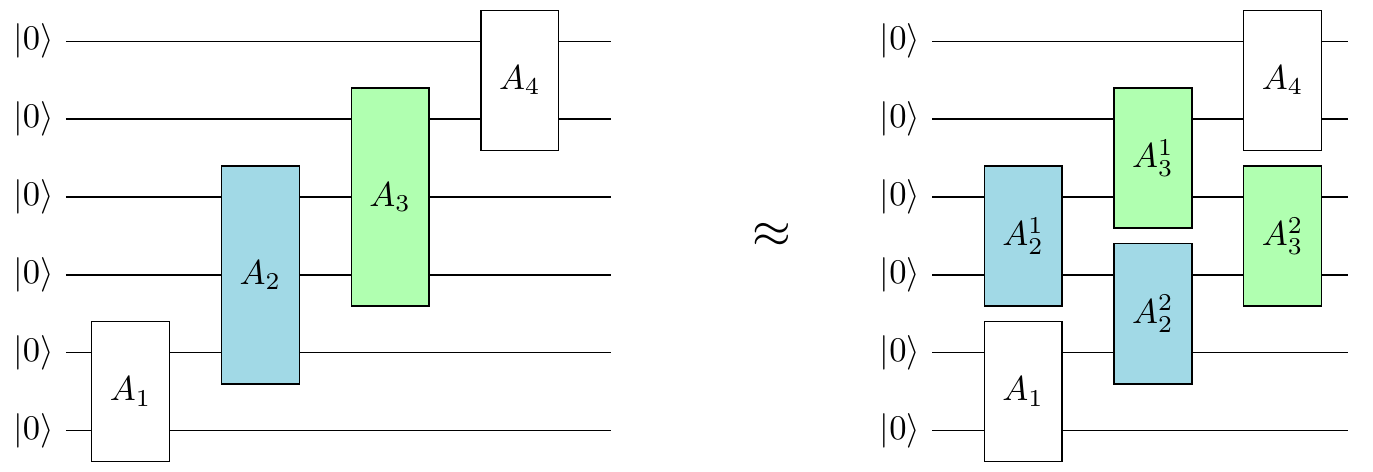}}
    \caption{\textbf{Approximate high bond dimension compilation.} a) The approximate decomposition of a 4-qubit unitary into a reverse stair case with 3 2-qubit unitaries.
    b) Using this decomposition is equivalent to initialising the diagonals adjacent to the main diagonal in a brick wall circuit. The circuit shown is a bond dimension 4 MPS which is decomposed into 2 diagonals in a brick wall circuit. Doubling the bond dimension adds another diagonal layer to the compilation.}
    \label{fig:ApproxComp}
\end{figure}

    An approximate compilation scheme has been proposed which has been demonstrated to effectively represent states of physical interest in tensor network simulations \cite{reverse_stair_ansatz}. Many-qubit unitary matrices are decomposed into a reverse stair-case of nearest neighbour unitary matrices, Fig.~\ref{fig:ApproxComp}a. When compiling an MPS into a circuit in this way, the diagonals adjacent to the central diagonal in a brick wall circuit are also initialised, and not set to the identity, Fig.~\ref{fig:ApproxComp}b. Training higher bond dimension MPS can then be viewed as a way to initialise more of the initial brick wall circuit. This compilation process is not exact, and hence errors will accumulate throughout this process and worsen the initialisation of the circuit.

    \section{Discussion}\label{sec:outlook}
    \subsection{Comparison to other methods}
    
    Below we compare conceptually similar ideas which have been proposed to overcome barren plateaus and accelerate the training of variational quantum circuits. This is not an exhaustive list but highlighting these examples helps highlight the benefits of MPS pre-training.
    
    \subsubsection{Warm-Start QAOA}
    The MPS initialisation method introduced here is one of many proposed methods to improve the training of PQCs. This solution is very similar in methodology to the warm-start QAOA algorithm \cite{warmStartQAOA}. Relaxations are applied to the optimization problem which make the problem efficiently solvable, and this is used to initialise the QAOA algorithm. The relative merits of these two methods ultimately depends on the relative effectiveness of relaxation methods and MPS based methods to approximate the optimal solution for optimization problems. There is currently no way to concretely answer this question, but there have been results demonstrating that tensor network based methods are competitive with state of the art commercial solvers for optimization problems \cite{PortfolioOptTN, QEnhancedOpt}. 

    \subsubsection{Layer-Wise Learning}
    Another procedure that has shown promising results is layer wise learning\cite{layer_wise_training, VQEvQAOA}. Layers in a quantum circuit are trained one at a time, keeping all other layers fixed. Finally layers are grouped together and trained simultaneously before training the entire circuit. The reasoning behind this method is very similar to that proposed here, to initialise the entire circuit with an approximation to the optimal solution, but instead of using MPS, the approximation is generated by sequential optimisation of layers in a PQC. Once again the question remains as to whether the layer wise trained approximation is better than MPS based approximations. Consider the fact that depth 1 or 2 nearest neighbour circuits, which are often considered in layer wise training regimes, can be faithfully represented with a bond dimension 2 MPS, and in fact this set of states is a restriction on the set of states that can be represented with a bond dimension 2 MPS. Layer wise training could be reformulated as sequential training of finite correlation length MPS, with the bond dimension of the restricted MPS growing with each trained layer. Seeing as the MPS initialisation scheme that we have introduced requires no restriction on the set of accessible states it should be the case that better initialisation states are accessible with the methods introduced here. It has been demonstrated that layer-wise learning suffers from abrupt transitions in trainability~\cite{layer_wise_learning_failure}; there exist circuit ansatz and cost functions where, below a threshold depth, piece wise training fails to minimize the cost function. 
    
    It has not escaped our attention that insights from tensor network optimisation techniques could be used to augment this initialisation method with a training scheme similar to layer wise learning. In the DMRG algorithm, the bond dimension of the MPS is gradually increased. A similar approach with brick wall circuits would involve sequentially training diagonals either side of the central diagonal. It remains to be seen if this training scheme could be as effective as layer wise learning in training a circuit combined with MPS pretraining. 

    \subsubsection{Entanglement Restriction}
    The authors of~\cite{entanglementMitigation} note that the impacts of vanishing gradients can be mitigated by restricting entanglement between hidden and visible qubits in a PQC. A qubit is visible if the output of that qubit is used to calculate a cost function, and it is hidden if it is ignored. They propose a number of schemes, including starting with no entanglement between hidden and visible nodes, having a fixed entanglement, and learning circuits which have low entanglement. The quantum circuit MPS formalism used in this work can easily be extended to the regime with hidden and visible qubits. In this case our initialisation scheme would resemble a fixed entanglement initialisation scheme, where the entanglement between the hidden and visible nodes is fixed by the bond dimension of the tensor network. They additionally propose including additional terms to the cost function, one to restrict entanglement growth between hidden and visible nodes, and one to add Langevin noise which acts to mitigate the effect of vanishing gradients. Both these schemes could be implementing on top of the initialisation scheme proposed here.

    \subsection{Alternative Tensor Network Structures}

    There are many tensor network geometries that are used to represent states with properties that are not effectively captured by MPS. Many of these have circuit counterparts which could be initialised in the same way. For example MERA networks~\cite{MERA} have been proposed as a basis for quantum machine learning 
    
\begin{figure}[H]
    {\centering
    \includegraphics[width=\linewidth]{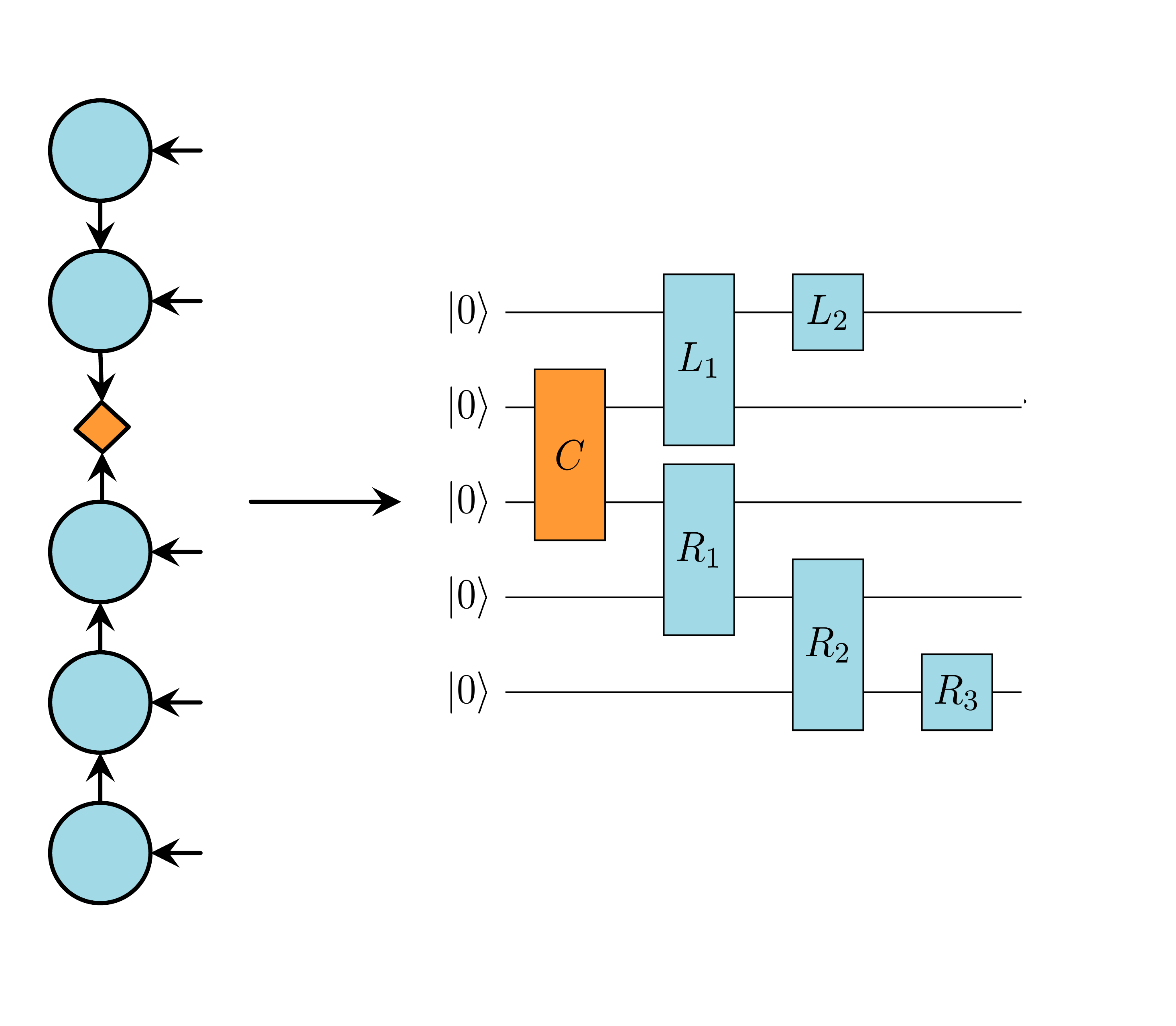}}

     \caption{\textbf{Gauge freedom and circuit depth.} The decomposition of a centre-gauge MPS into a quantum circuit. This circuit has a lower depth than left or right canonical MPS, used throughout this work. This may additionally mitigate the effect of vanishing gradients. Tensors are isometries from the space of incoming arrows to the space of outgoing arrows. For bond dimension 2 there exist low depth circuits that can exactly represent the central tensor, $C$~\cite{QMPS}. For higher bond dimension MPS the central tensor would need to be approximately compiled.}
    \label{fig:TNCircuit}
\end{figure}

    MPS have a freedom in the location of the \emph{othogonality centre} when put in canonical form. In all results above, the MPS are put into left or right canonical form. The circuits to represent these states have a depth at least as large as the number of qubits. However choosing a \emph{mixed canonical} form actually reduces the circuit depth needed to initialise the MPS, Fig.~\ref{fig:TNCircuit}. A circuit initialised in this way may have larger initial gradients, as gradients vanish as a function of circuit depth.
    
    \subsection{Consequences for tensor network simulations}
    
    It is interesting to note that the bond dimension of the MPS needed to represent the circuit increases as training proceeds. Deep brick wall circuits represent a restricted class of high bond dimension tensor networks. 
    Ordinarily any variational calculations with these extremely large bond dimension tensor networks would be impractical on quantum devices because of the difficulty in optimizing these circuits. To simulate spin systems with high bond dimension tensor networks it would be possible to simulate up to the classically feasible limit, translate the tensor network into circuits, and then seed a higher bond dimension simulation with the classical tensor network as is done here. This could open up the possibility of very large bond dimension tensor network simulations of spin systems on NISQ devices.

\section{Acknowledgements And Contributions}

JD, FB, and AGG were supported by the EPSRC through grants EP/L015242/1, EP/L015854/1 and
EP/S005021/1. This work is supported by the Engineering and Physical Sciences Research Council grant number EP/L015242/1. VM is supported by ESPRC Prosperity Partnership grant EP/S516090/1. LW and FB are supported by the EPSRC CDT in Cross-Disciplinary Approaches to Non-Equilibrium Systems (CANES) via grant number EP/L015854/1. The project was conceived in group discussions. JD and FB wrote code to translate MPS into circuits, and to continue training. VM wrote classical MPS training methods inspired by DMRG. The manuscript was written by JD and FB.  

\end{multicols}
\newpage

\printbibliography
\end{document}